# θ-Quantization of the Orbital Angular Momentum Operator by means of Noncommutative Geometry

Takeo Miura

Advanced Institute for Mathematical Science

1-24-4 Motojyuku, Higashimatuyama, Saitama 355-0063, Japan

E-mail:t_miura@msg.biglobe.ne.jp

## Abstract

The eigenfunctions and eigenvalues of orbital angular momentum operator on noncommutative lattice for a circle poset by θ-quantization are constructed, and it is demonstrated that they are equivalent to those of the conventional quantum mechanics.

## I Introduction

We construct the eigenfunctions and their eigenvalues of orbital angular momentum(OAM) operator on noncommutative lattices for the circle poset. We have already studied on the possible construction of nonlocal field of a quantum mechanical system such as the θ-quantization of the wave function for the Aharonov-Bohm effect modular momentum in Reference [1]. At present we are in the midst of investigating the nonlocality of angular variable on an orbit with quantization of canonically conjugate pair, angle and orbital angular momentum. Above all, we show in this paper that the eigenfunctions and eigenvalues of orbital angular momentum operator on noncommutative lattices for a circle poset by θ-quantization are equivalent to those of the conventional quantum mechanics.

## II θ − quantization of OAM operator on the circle poset

We construct the $\theta$-quantization of a particle on a noncommutative lattice for the circle poset.

At first, we describe the quantization of the canonically conjugate pair, angle and orbital angular momentum on the circle. The angle is a multivalued or discontinuous variable on the corresponding phase space. Therefore, we replace the angle $\varphi$ by the smooth periodic functions $\cos\varphi$ and $\sin\varphi$. In the case of the canonical pair $(\varphi, P_\varphi)$, $P_\varphi$: orbital angular momentum(OAM), the phase space $S_{\varphi,P_\varphi} = \{\varphi \in \mathcal{R} \mod 2\pi, P_\varphi \in \mathcal{R}\}$ has the global topological structure $S^1 \times \mathcal{R}$ of a cylinder on which the Poisson brackets of the three functions $\cos\varphi$, $\sin\varphi$ and $P_\varphi$ obey the Lie algebra of the Euclidean group E(2) in the plane. Then we have the basis for the quantization of the system in terms of irreducible unitary representations of the group E(2) or those of its covering groups [2].

Equivalently, the real line $R^1$ is the universal covering space of the circle, and the fundamental group $\pi_1(S^1) = \mathbb{Z}$ acts on $R^1$ by translation $R^1 \ni x \to x + N$, $N \in \mathbb{Z}$. The quotient space of this action is $S^1$ and the projection : $R^1 \to S^1$ is given by $R^1 \ni x \to e^{i2\pi x} \in S^1$. The domain of a typical Hamiltonian for a particle on $S^1$ can be obtained from functions $\psi_\theta$ on $R^1$ transforming under an irreducible representation of $\pi_1(S^1)=\mathbb{Z}$, $N \to N + \theta$, $\rho_\theta : N \to e^{iN\theta}$ according to $\psi_\theta(x + N)=e^{iN\theta}\psi_\theta(x)$. Thus θ is introduced [3].

Due to the fact that the subgroup SO(2) $\cong$ $S^1$ is multiply connected, these representations allow for the functional OAM $\ell=\hbar(N + \theta)$, $N \in \mathbb{Z}$, $\theta \in [0,1)$.
In any irreducible unitary representations of E(2) or any of its covering groups, the generator L of the rotations has arbitrarily large positive and negative eigenvalues. These representations may all be implemented in a Hilbert space $L^2(S^1, d\varphi/2\pi)$ of functions $\psi(\varphi)$ with the scalar product

$$(\psi_2,\psi_1) = \int_0^{2\pi} \frac{d\varphi}{2\pi} \psi_2^*(\varphi)\psi_1(\varphi). \tag{1}$$

One obtains a different quantization, called θ-quantization, for each choice of the parameter θ in the irreducible unitary representations of the different covering group $\widetilde{E}(2)$, universal covering group of E(2).

We can define the following generators (see Reference [2]).

$$\frac{1}{\hbar}L_\theta = \widetilde{L_\theta} = \frac{1}{i}\partial_\varphi + \theta, \quad X_1 = r\cos\varphi, \quad X_2 = r\sin\varphi. \tag{2}$$

The Hilbert space $L^2(S^1, d\varphi/2\pi)$ with the scalar product (1) has the orthonormal basis

$$e_n(\phi)=e^{in\varphi}, \quad n \in \mathbb{Z}. \tag{3}$$

The functions (3) are eigenfunctions of OAM-operator $L_\theta$ ;

$$L_\theta\, e_n = \hbar(n + \theta)e_n. \tag{4}$$

The different θ leads to different spectra of $L_\theta$ and therefore such operators are not unitarily equivalent. The different irreducible unitary representations corresponding to different θ are all realized on the same Hilbert space with the basis (3). By making the unitary transformations

$$e_n(\varphi)=e^{in\varphi} \to e_{n,\theta}(\varphi) = e^{i\theta\varphi}e_n(\varphi) = e^{i(n+\theta)\varphi} \quad \forall n \in \mathbb{Z}, \tag{5}$$

we can define a separate Hilbert space $L^2(S^1, d\varphi/2\pi, \theta)$ for each θ. In these Hilbert spaces the generators (2) now have the common form

$$\frac{1}{\hbar}L_\theta=\frac{1}{i}\partial_\varphi, \quad X_1 = r\cos\varphi, \quad X_2 = r\sin\varphi. \tag{6}$$

That is, the operators are now independent of θ, the dependence of which is shifted to the basis(5).

In the following sections, we must consider the representation of the cyclic group of order N. Let G = $\mathbb{Z}_N$ be the cyclic group of order N acting by translations on X = $\{1, 2, \cdots, N\}$. The crossed product algebra $\mathbb{C}(X) \rtimes G$ is the algebra generated by elements U and V subject to the relations

$$U^N = 1, \quad V^N = 1, \quad UVU^{-1} = \lambda V, \quad \text{where } \lambda = e^{2\pi i/N}.$$

Here, U is a generator of $\widehat{\mathbb{Z}_N}$ ( the dual group of $\mathbb{Z}_N$) and we have used the isomorphism $\mathbb{C}(X) \simeq \mathbb{C}\widehat{\mathbb{Z}_N}$, where an isomorphism $\mathbb{C}(X) \rtimes \widehat{\mathbb{Z}_N} \simeq M_N(\mathbb{C})$. ( See, Serre and Swan theorem .)

In the case of the above Hilbert space $L^2(S^1, d\varphi/2\pi, \theta)$, the orthonormal basis(5) and the eigenvalues (4) of OAM-operator are replaced by

$$e_{n,\theta}(\phi) = e^{i(n+\theta)\varphi/N}, \tag{7}$$

and

$$L_\theta e_{n,\theta} = \hbar((n+\theta)/N) e_{n,\theta}. \tag{8}$$

Ⅲ  Line Bundles on Circle poset and $\theta$-quantization (see References [3], [4])

We consider the $\theta$-quantization of operators and eigenfunctions for a noncommutative lattice on the circle in the Hilbert space $L^2(S^1, d\varphi/2\pi)$.

One constructs the algebraic analogue of the trivial bundle on the lattice (the circle poset) with a gauge connection so that the corresponding orbital angular momentum operator has an approximate spectrum and their eigen functions.

The algebra A associated with any noncommutative lattice of the circle can indeed be approximated by algebras of matrices, as it is AF(approximately finite dimensional). The simplest approximation is just a commutative algebra $\mathbb{C}(A)$ of the form

$$\mathbb{C}(A) \simeq \mathbb{C}^N = \{c = (\lambda_1, \lambda_2, \cdots, \lambda_N)\}, \quad \lambda_i \in \mathbb{C}\}. \tag{9}$$

The algebra(8) can produce a noncommutative lattice with 2N points by considering a particular class of not necessarily irreducible representations as in Fig. 11.1. in Reference[3]. In this Figure, the top points correspond to the irreducible one dimensional representations

$$\pi_i : \mathbb{C}(A) \to \mathbb{C}, \quad c \mapsto \pi_i(c) = \lambda_i, \quad i = 1, \cdots, N.$$

As for the bottom points, they correspond to the reducible two dimensional representations

$$\pi_{i+N} : \mathbb{C}(A) \to M_2(\mathbb{C}), \quad c \mapsto \pi_{i+N}(c) = \begin{pmatrix} \lambda_i & 0 \\ 0 & \lambda_{i+1} \end{pmatrix}$$

with the additional condition that N+1=1.

The space $M_2(\mathbb{C})$ we are dealing with has two parts a and b. Thus the algebra A is just the direct sum $\mathbb{C} \oplus \mathbb{C}$ of the copies of $\mathbb{C}$. An element $f \in A$ gives two complex numbers $f(a), f(b) \in \mathbb{C}$.

Let $(\mathcal{H}, D)$ be a 0-dimensional K-cycle over A (using, for example, the Dixmier trace), then $\mathcal{H}$ is finite-dimensional and the representation of A in $\mathcal{H}$ corresponds to a decomposition of $\mathcal{H}$ as a direct sum $\mathcal{H} = \mathcal{H}_a \oplus \mathcal{H}_b$, with the action of A given by

$$f \in A \mapsto \begin{bmatrix} f(a) & 0 \\ 0 & f(b) \end{bmatrix}.$$

If we write D as a 2×2 matrix in the decomposition,

$$D = \begin{bmatrix} D_{aa} & D_{ab} \\ D_{ba} & D_{bb} \end{bmatrix},$$

we can ignore the diagonal elements since they commute exactly with the action of A. We take $D_{ba} = D_{ab}^*$ and $D_{ba}$ is a linear mapping from $\mathcal{H}_a$ to $\mathcal{H}_b$. We denote this linear mapping by M,

$$D = \begin{bmatrix} 0 & M^* \\ M & 0 \end{bmatrix}.$$

The partial order is determined by the inclusion of the corresponding kernels. A better approximation is obtained by approximating compact operator, with finite dimensional matrices of increasing rank.

The finite projective module of sections $\mathcal{E}$ associated with the trivial line bundle is just $\mathbb{C}(A)$ itself :

$$\mathcal{E} = \mathbb{C}^N \approx \{ \eta = (\mu_1, \mu_2, \cdots, \mu_N), \mu_i \in \mathbb{C} \}.$$

The action of $\mathbb{C}(A)$ on $\mathcal{E}$ is simply given by

$$\mathcal{E} \times \mathbb{C}(A) \to \mathcal{E}, \quad (\eta, c) \mapsto \eta c = (\eta_1 \lambda_1, \eta_2 \lambda_2, \cdots, \eta_N \lambda_N).$$

On $\mathcal{E}$ there is a $\mathbb{C}(A)$-valued Hamiltonian structure $\langle \cdot, \cdot \rangle$ such that

$$\langle \eta', \eta \rangle = (\eta_1'^* \eta_1, \eta_2'^* \eta_2, \cdots, \eta_N'^* \eta_N) \in \mathbb{C}(A). \tag{10}$$

It is necessary that we have a K-cycle $(\mathcal{H}, D)$ over $\mathbb{C}(A)$. We take $\mathbb{C}^N$ for $\mathcal{H}$ on which we represent elements of $\mathbb{C}(A)$ as a diagonal matrices c,

$$( c \in \mathbb{C}(A) ) \mapsto \mathrm{diag}(\lambda_1, \lambda_2, \cdots, \lambda_N) \in \mathcal{B}(\mathbb{C}^N) \simeq M_N(\mathbb{C}).$$

Now our triple $(\mathbb{C}(A), \mathcal{H}, D)$ should be zero dimensional, the $\mathbb{C}$-valued scalar product associated with the Hamiltonian structure (9) should be

$$\langle \eta', \eta \rangle = \sum_{j=1}^{N} \eta_j'^* \eta_j = \mathrm{tr}\langle \eta', \eta \rangle, \qquad \eta', \eta \in \mathcal{E}.$$

We then consider $M_N(\mathbb{C})$ algebra. By identifying $N + j = j$, we take for the operator D, the N×N self-adjoint matrix with elements

$$D_{ij} = \frac{1}{\varepsilon\sqrt{2}}(M^*\delta_{i+1,j} + M\delta_{i,j+1}), \qquad i,j = 1,\cdots,N. \tag{11}$$

where M is any complex number of modulus one : $MM^* = 1$.

Let $A = C^\infty(S^1)$ denote the algebra of smooth complex valued functions on the circle. The fundamental class of $S^1$ in de Rham cohomology has the class of the differential 1-form $\omega = z^{-1}dz$.

As for the connection 1-form $\rho$ on the bundle $\mathcal{E}$, we take it to be the Hermitian matrix with elements

$$\rho_{ij} = \frac{1}{\varepsilon\sqrt{2}}(\sigma^* M^*\delta_{i+1,j} + \sigma M\delta_{i,j+1}) \tag{12}$$

$$\sigma = e^{-i\theta\varphi/N} - 1, \qquad i,j = 1,\cdots,N. \tag{13}$$

We find the above expression(13) of $\sigma$ in the following way.

If $M \neq 0$ then the representation $\pi : \Omega^*(A) \to \mathcal{L}(\mathcal{H})$ is injective on $\Omega^1(A) \approx \Omega_D^1(A)$ where $\Omega_D^1$ is Conne's Differential 1-form, $\Omega_D^1(A) \approx \pi(\Omega_1 A)$ and this space coincides with the A-bimodule of bounded operator on $\mathcal{H}$ of the form $\omega_1 = \sum_j a_0^j [D, a_1^j]$, $a_i^j \in A$. We have (Reference [4])

$$\pi(\lambda\, ede + \mu(1-e)de ) = \begin{bmatrix} 0 & -\lambda M^* \\ \mu M & 0 \end{bmatrix} \in \mathcal{L}(\mathcal{H}).$$

A vector potential V is given by a self-adjoint element of $\Omega_D^1$, i.e., by single complex number $\Phi$, with

$$\pi(V) = \begin{bmatrix} 0 & \Phi M^* \\ \Phi M & 0 \end{bmatrix}.$$

Since $V = -\Phi ede + \Phi(1-e)de$, Its curvature is

$$\Theta = dV + V^2 = -\bar{\Phi}dede - \Phi dede + (\bar{\Phi}ede - \Phi(1-e)de),$$

and, using the equalities $ede(1-e) = ede$, $e(de)e = 0$, $(1-e)de(1-d) = 0$, we have

$$\Theta = dV + V^2 = -(\Phi + \bar{\Phi})dede - (\Phi\bar{\Phi})dede. \tag{14}$$

Under the representation $\pi$, we have $\pi(de) = \begin{bmatrix} 0 & -M^* \\ M & 0 \end{bmatrix}$ and $(dede) = \begin{bmatrix} -M^*M & 0 \\ 0 & -MM^* \end{bmatrix}$.

This yields the formula for the Yang=Mills action,

$$YM(V) = 2(|\Phi+1|^2-1)^2 \text{Trace}((M^*M)^2),$$

where $\Phi$ is a arbitrary complex number. Under the minimum extrema of the YM(V)=0, this reduces to $(|\Phi+1|^2 - 1)^2 = 0$. Thus, we obtain $\Phi = \sigma = e^{-i\theta\varphi/N} - 1$.

Also we put this $\Phi$ into (14), we obtain $\Theta = 0$, which implies $dV + V^2 = 0$, thus the curvature of $\rho$ vanishes, that is $d\rho + \rho^2 = 0$.

IV  The representation of the eigenfunctions and eigenvalues of OAM-operator on the
    circle poset in Hilbert space

If $c = \text{diag}(\lambda_1, \lambda_2, \cdots, \lambda_N)$, then any such c is given by

$\lambda_1 = \lambda$, $\lambda_2 = e^{i\varphi/N}\lambda$, $\cdots$, $\lambda_m = e^{i\varphi(m-1)/N}\lambda$, $\cdots$, $\lambda_N = e^{i\varphi(N-1)/N}\lambda$, with $\lambda$ not equal to 0.
The covariant derivative $\nabla_\varphi$ on $\mathscr{E}$ : $\mathscr{E} \to \mathscr{E} \otimes_{c(A)} \Omega^1(\mathbb{C}(A))$ is given by

$$\nabla_\varphi \eta = [D, \eta] + \rho\eta, \qquad \forall\, \eta \in \mathscr{E}.$$

Components of (11), (12) with $\eta_j$ are as follows.

$$D\eta_j = \frac{1}{\sqrt{2\varepsilon}}(m^*\delta_{i+1,j} + m\delta_{i,j+1})\eta_j$$

$$= \frac{1}{\sqrt{2\varepsilon}}(m^*\eta_{j-1} + m\eta_{j+1}), \qquad j = 1, 2, \cdots, N.$$

$$\rho\eta_j = \frac{1}{\sqrt{2\varepsilon}}(m^*\sigma^*\delta_{i+1,j} + m\sigma\delta_{i,j+1})\eta_j$$

$$= \frac{1}{\sqrt{2\varepsilon}}(m^*\sigma^*\eta_{j-1} + m\sigma\eta_{j+1}). \qquad j = 1, 2, \cdots, N.$$

Here, we consider the transition ($j \to j-1$) of component state for each j, therefore D is decomposed into $D = D^+ + D^-$, and as operators we take $D^-$ and $\rho^-$. We can set $m = m^* = 1$. Therefore

$$D^-\eta_j = \frac{1}{\sqrt{2\varepsilon}}\eta_{j-1},$$

$$\rho^-\eta_j = \frac{1}{\sqrt{2\varepsilon}}\sigma^*\eta_{j-1}.$$

Thus

$$\nabla_\theta^-\eta_j = \frac{1}{\sqrt{2\varepsilon}}(1 + \sigma^*)\eta_{j-1}$$

$$= \frac{1}{\sqrt{2\varepsilon}}e^{i\theta\varphi/N}\eta_{j-1}.$$

Here, we can take $\lambda = e^{i\varphi/N}e^{i\theta\varphi/N}$, thus $\eta_j = e^{i(j+\theta)\varphi/N}$, and obtain the eigenfunctions operator of OAM on circle poset,

$$e_{j,\theta} = \eta_j = e^{i(j+\theta)\varphi/N}, \qquad j = 1, \cdots, N,$$

which is the same as (7).

Because eigenvalues equation is the following

$$L_\theta\eta_j = L_\theta e_{j,\theta} = \hbar\left(\frac{1}{i}\partial_\varphi\right)e^{i(j+\theta)\varphi/N}$$

$$= (\hbar((j+\theta)/N))\eta_j,$$

we obtain the eigenvalues $\hbar(j+\theta)/N$, which is the same as (8).

V Conclusion

We demonstrated that the eigenfunctions and eigenvalues of OAM-operator on noncommutative lattices for a circle poset by θ-quantization are equivalent to those of the conventional quantum mechanics. Considering these results, we advocate that the concept of nonlocality with quantum fields is crucially important.